\documentclass[journal=apchd5,manuscript=article]{achemso}

\usepackage[version=3]{mhchem} 
\usepackage[T1]{fontenc}       
\usepackage{amsmath,amsthm,amssymb,amsfonts,nccmath}
\usepackage{graphicx,color,soul}
\usepackage{epstopdf}
\DeclareGraphicsExtensions{.pdf}
\usepackage{color}
\usepackage[utf8]{inputenc}

\usepackage{ulem}

\author{Michelle L. Solomon}
\affiliation{Materials Science and Engineering, Stanford University, 496 Lomita Mall, Stanford, California 94305, USA}
\author{John M. Abendroth}
\affiliation{Materials Science and Engineering, Stanford University, 496 Lomita Mall, Stanford, California 94305, USA}
\author{Lisa V. Poulikakos}
\affiliation{Materials Science and Engineering, Stanford University, 496 Lomita Mall, Stanford, California 94305, USA}
\altaffiliation{Mechanical and Aerospace Engineering, University of California San Diego, 9500 Gilman Drive, La Jolla, California 92093, USA}
\author{Jack Hu}
\affiliation{Materials Science and Engineering, Stanford University, 496 Lomita Mall, Stanford, California 94305, USA}
\author{Jennifer A. Dionne}
\affiliation{Materials Science and Engineering, Stanford University, 496 Lomita Mall, Stanford, California 94305, USA}
\email{msolomo8@stanford.edu, jdionne@stanford.edu}
\title{Fluorescence-Detected Circular Dichroism of a Chiral Molecular Monolayer with Dielectric Metasurfaces}

\begin{tocentry} 
\includegraphics{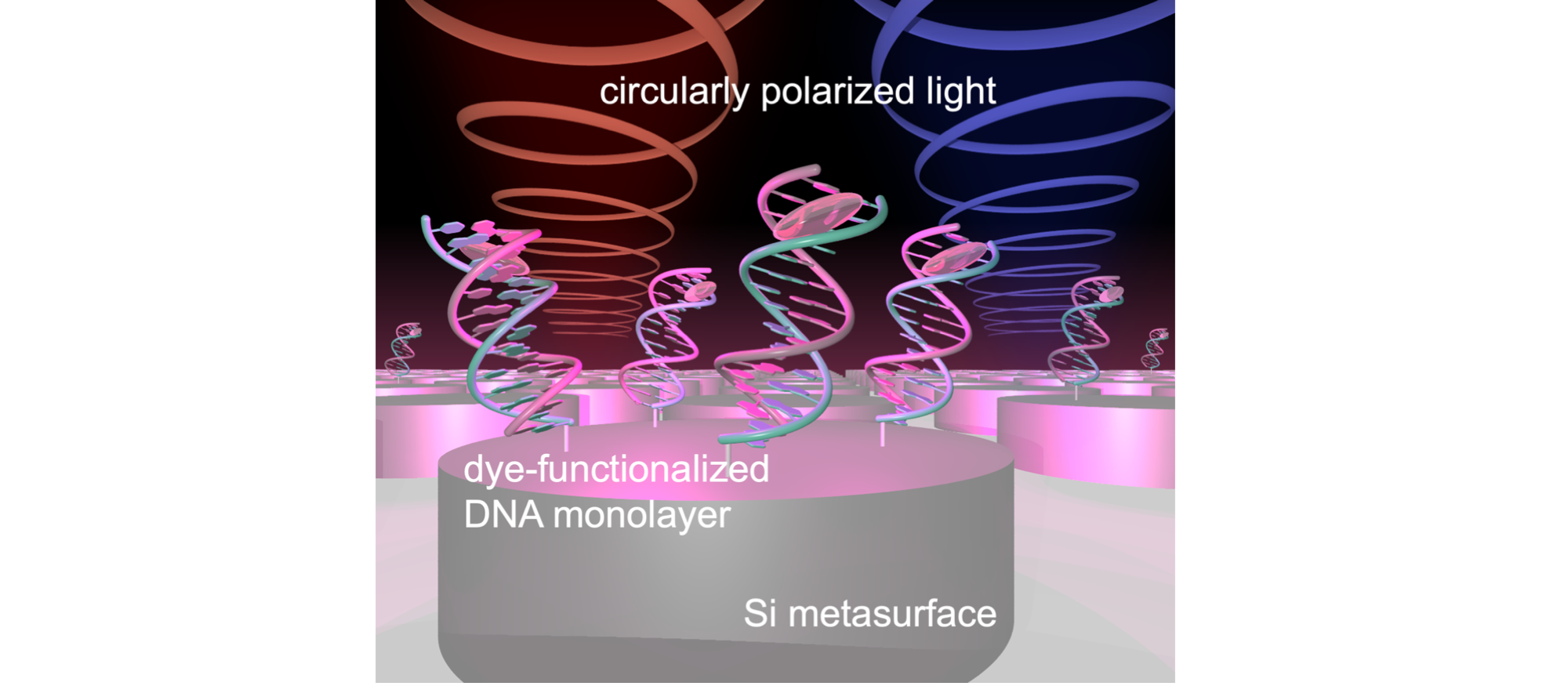}
\end{tocentry}

\begin{document}

\begin{abstract}
Strong enhancement of molecular circular dichroism has the potential to enable efficient asymmetric photolysis, a method of chiral separation that has conventionally been impeded by insufficient yield and low enantiomeric excess. Here, we study experimentally how predicted enhancements in optical chirality density near resonant silicon nanodisks boost circular dichroism. We use fluorescence-detected circular dichroism spectroscopy to measure indirectly the differential absorption of circularly polarized light by a monolayer of optically active molecules functionalized to silicon nanodisk arrays. Importantly, the molecules and nanodisk antennas have spectrally-coincident resonances, and our fluorescence technique allows us to deconvolute absorption in the nanodisks from the molecules. We find that enhanced fluorescence-detected circular dichroism signals depend on nanophotonic resonances in good agreement with simulated differential absorption and optical chirality density, while no signal is detected from molecules adsorbed on featureless silicon surfaces. These results verify the potential of nanophotonic platforms to be used for asymmetric photolysis with lower energy requirements.    
\end{abstract}

Chirality, or handedness, is a fundamental property of all living organisms, from biological building blocks such as DNA and amino acids to macroscopic structures. Chirality also features prominently in many synthetic molecules, with over 50\% of pharmaceuticals and 40\% of agrochemicals existing  as enantiopure forms or racemic mixtures.\cite{nguyen2006chiral,jeschke2018current} These enantiomers can have distinct efficacy in biological systems, making the ability to distinguish mirror-image molecules with high sensitivity and maximize enantiomeric excess (e.e.) in asymmetric synthesis crucial tasks. Circular dichroism (CD), defined as the selective absorption of circularly polarized light (CPL), is commonly used to differentiate enantiomers via CD spectroscopy, though it typically requires relatively high sample concentrations or long optical path lengths.\cite{kelly2005study,zhao2017chirality} This differential absorption between left- and right-CPL, $\text{A}^\text{L}-\text{A}^\text{R}$ ($\Delta\text{A}$), has also inspired efforts to use CPL as a reagent in enantioselective synthesis or for photolysis as early as 1929.\cite{kuhn1929photochemische,feringa1978biomimetic} However, due to the low differential absorption cross section of molecules, the yield \& e.e. achieved by decomposition of optical isomers with light alone falls below industrially relevant state-of-the-art techniques to maximize enantiopurity.\cite{inoue1992asymmetric,flores1977asymmetric,balavoine1974preparation}

A variety of methods have sought to improve sensitivity of chiral differentiation in CD spectroscopy. For example, nonlinear spectroscopies utilizing second harmonic generation that are sensitive to surfaces and interfaces\cite{lin2014situ,burke2003experimental,petralli1993circular} and single molecule spectroscopy have enabled enantiomeric detection at the monolayer to few-to-single molecule regime.\cite{hassey2009dissymmetries,cyphersmith2012chiroptical} To expand upon these spectroscopic techniques, superchiral electromagnetic fields arising from the interference of chiral plane waves have been shown to increase dissymmetry in chiral excitation, but the position of the superchiral fields within the resulting standing wave nodes limits utility.\cite{cohen:2011,cohen:2010} Recently, nanophotonic architectures have received considerable attention due to their potential to manipulate chiral evanescent near fields while maintaining high field strength.\cite{gansel2009gold,Poulikakos:2018,esposito:2015,Ferry:2015,naliu:2015,liu:2018,kuzyk2012dna,govorov:2010,wang2015circular,zhao2012twisted} These near fields have been predicted to enhance the enantioselective rates of molecular absorption and can be achieved using plasmonic
\cite{schaferling2014helical,giessen:2012physrevx,hendry:2010,hendry:2012,kneer2018circular,garcia2018enantiomer,tullius2015superchiral,nesterov2016role}and high-refractive-index nanostructures. \cite{aitzol,shings,solomon2018enantiospecific,hu2020high,mohammadi,ben2013chirality,zhang:2017} In experiments, many of these approaches enable highly effective enantiomeric sensors, but often exhibit negligible spectral overlap between chiral molecular absorption and nanoantenna resonances, which is necessary to acheive enantioselective photolysis with high yield and e.e.\cite{kagan:1974,shings,solomon2018enantiospecific} Therefore, rather than enhancing differential absorption rates by the molecules themselves, resulting CD signals in the visible and near-IR are mainly due to intrinsic (extrinsic) chirality of 3D (2D) nanostructures, or induced CD in lossy platforms.\cite{lee2017microscopic}  Furthermore, these techniques struggle to distinguish molecular absorption from total absorption making it difficult to unveil and optimize the near-field mechanism behind chiral-optical enhancements in molecules.\cite{garcia2019enhanced,mohammadi2019accessible}

Here, we demonstrate enhanced enantioselective absorption in chiral molecular monolayers using nanostructures with optical resonances spectrally matched to the molecular CD. Sub-wavelength, periodic arrays of silicon disks (hereafter, “metasurfaces”) are functionalized with self-assembled monolayers of fluorescently-labeled oligonucleotide strands with visible-frequency CD (see Figure 1a), attributed to dye binding within the helical DNA environment. We use fluorescence-detected circular dichroism (FDCD) to perform a background-free measurement, as the dye fluorescence is distinguishable from that in the metasurface or substrate. While negligible FDCD is observed on unpatterned films, we observe strong, red-shifting FDCD on disks with increasing radius, in agreement with calculations. We show that our method can distinguish conformation in molecular monolayers, which we validate with in-situ measurements of FDCD sign reversal during DNA dehybridization. Our results exhibit enhancement of intrinsic molecular CD, en-route to enantioselective photolysis.   

\begin{figure}
\includegraphics{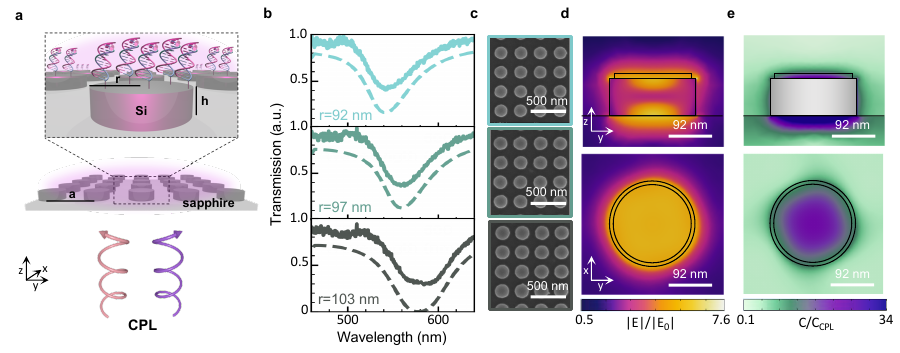} 
\caption{\label{fig1} a) Schematic: dye/DNA functionalized to metasurface b) Array transmission: radius, $r$=92-103nm, height, $h$=80nm, pitch, $a$=300nm c) Array SEMs d) Electric field enhancement on resonance, $\lambda$=570nm. e) Maximized C enhancement, $\lambda$=565nm. d\&e: $r$=92nm metasurface; top through disk center; bottom 5nm above disk/layer interface. Simulated chiral layer radius is $r_l$=82nm, avoiding mesh artifacts at disk edge, with layer height, $h_l$=10nm.}
\end{figure}

Metasurfaces were engineered to have concurrent electric and magnetic dipolar Mie resonances coinciding with the molecular monolayer absorption, a condition known as the first Kerker condition at which optical chirality density was found to be highly enhanced in our previous work.\cite{solomon2018enantiospecific} These overlapping modes enhance optical chirality density due to strong electric and magnetic near fields that maintain the phase properties of incident CPL, and disk nanoantennas enable facile tuning of such resonances (see SI and Figure S1 for further detail).\cite{solomon2018enantiospecific,hu2020high,mohammadi2019accessible,graf2019achiral,kivshar:2013, siday2019terahertz,yang2018nonradiating,decker2015high,decker2016resonant} Nanodisk arrays were fabricated in single-crystalline silicon layers grown on sapphire substrates. The fabricated arrays are $100\mu\text{m}\times100\mu\text{m}$, with nominal disk height $h$=80nm, disk radii $r=$90-105nm, and pitch $a$=300nm. Figure 1b includes experimental transmission spectra of three example bare silicon metasurfaces with nominal disk radii of $r$=92, 97, and 103nm, with representative SEMs in Figure 1c. We simulate these metasurfaces with a layer of chiral medium on top, where layer height $h_l$=10nm, and layer radius, $r_l=r$-10nm (see SI), finding that the simulated transmission (dotted line, Figure 1b) is in good agreement with experiments. Dips in the transmission spectra correspond to concurrent electric and magnetic Mie resonances; the simulated electric field plots of Figure 1d are indicative of these overlapping resonances and show strong electric fields extending into the chiral layer.\cite{solomon2018enantiospecific} Near these resonances, and just blue-shifted ($\lambda$=565nm vs. $\lambda$=570nm), optical chirality density in the chiral layer reaches a maximum enhancement (Figure 1e). Here, we define the optical chirality density as $\text{C}=-\frac{\omega}{2}\text{Im}(\textbf{D}^*\cdot\textbf{B}$) and the enhancement factor as C/$\text{C}_\text{CPL}$, where $\text{C}_\text{CPL}=\frac{\epsilon_{0}\omega}{2c}E_{0}^2$, the optical chirality density of CPL alone.\cite{cohen:2010,barron:2009, Poulikakos:2018} The small shift between peak C/C$_{\text{CPL}}$ and resonance center wavelength occurs due to the balance between the relative phases and intensities of the electric \& magnetic fields.\cite{shings}
\begin{figure}
\includegraphics{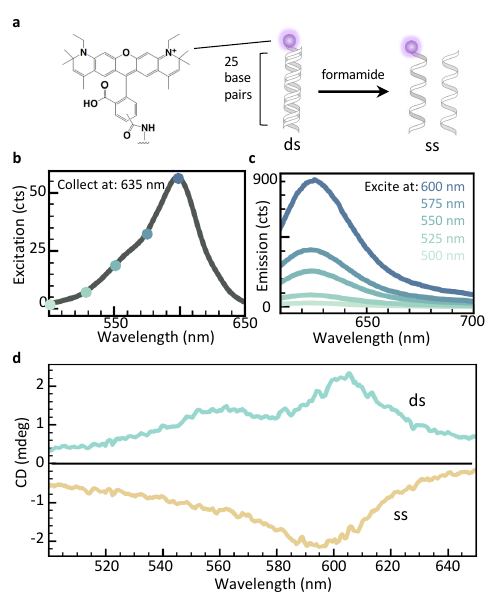} 
\caption{\label{fig2} a) Structure of $\text{Atto}^{\text{TM}}$590 dye functionalized to 5'-end of DNA \& de-hybridization scheme b) Excitation spectrum of dye-dsDNA in PBS c) Emission spectra of dye-dsDNA in PBS d) CD of dye-dsDNA \& dye-ssDNA in PBS, 30$\mu$M.}
\end{figure}

The fluorophore, $\text{Atto}^{\text{TM}}$590, was covalently attached to the 5’-end of a 25 base pair DNA sequence with a thiol modification on the 3’-end (Figure 2a, see SI for details). The complex absorbs strongly from 500-650nm, with a peak in excitation at 600nm and an emission maximum at 615nm that depends on the excitation wavelength (Figure 2b,c). When hybridized with its fully complementary strand, the dye-dsDNA complex exhibits a positive CD signal in its visible absorption band.\cite{kringle2018temperature,norden1982structure,neelakandan2014ground,tuite2018linear} Interestingly, the sign of the CD signal is reversed when the DNA molecules are denatured as the dye molecules experience the different local environment of the single strand's secondary structure (Figures 2d, S2).\cite{stokes2007making} This enantiomer-reversal-like behavior is particularly useful to probe changes in the sign of the CD signal on the same substrates without varying surface density.

The metasurfaces were functionalized with self-assembled monolayers of the dye-DNA complexes with widely-used silane chemistry (Figure 3a).\cite{nakatsuka2018aptamer,cheung2020detecting} Terminal amine groups on vapor-deposited films of (3-aminopropyl)trimethoxysilane were crosslinked to pre-hybridized thiol-modified DNA strands using \textit{m}-maleimidobenzoyl-\textit{N}-hydroxysuccinimide ester (See SI). Monolayers of DNA prepared by silanization on Si/SiO$_2$ typically feature surface densities of $\sim$10$^{12}$ molecules/cm$^2$, which represents an approximate upper bound to the limit of detection of our technique.\cite{strother2000synthesis, pirrung2002make} To validate surface functionalization, metasurface fluorescence spectra were collected both before and after DNA assembly. Strong fluorescence near the long-pass edge at 600nm can be attributed to background fluorescence from the sapphire substrate. However, a significant increase in emission intensity from 610-650nm following functionalization confirms that fluorescence from the monolayer is distinguishable from that of the substrate (Figure 3b). 

\begin{figure}
\includegraphics{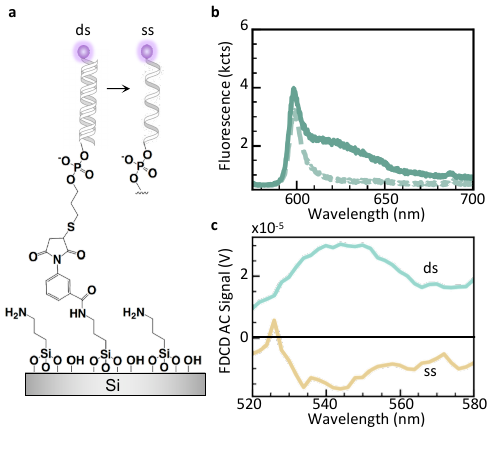}
\caption{\label{fig3}a) Functionalization scheme and de-hybridization on substrate b) Fluorescence of dye-functionalized 92-nm-radius array (solid) vs dye-less (dotted) c) FDCD sign reversal from dye-dsDNA to dye-ssDNA}.
\end{figure}

To detect the CD of the monolayers indirectly using fluorescence, we built a table-top polarization-sensitive spectrometer that performs a lock-in measurement to collect excitation spectra (see SI). For this measurement, we use a 635nm bandpass filter that transmits the fluorescence of the dye alone. First, we studied the FDCD signal dependence on the hybridization state of the DNA monolayers which, from solution/ensemble-CD measurements, reverses sign between double- and single-stranded forms. Monolayer-functionalized metasurfaces were mounted within a cuvette of 1x phosphate buffered saline (PBS) to maintain the helical tertiary structure of dsDNA when tethered to surfaces during measurements (Figure S3). Using a metasurface of $r$=92nm functionalized with dye-dsDNA complexes, a positive FDCD signal is measured from 520-580nm (Figure 3c). Then, PBS was removed and replaced with formamide to lower the DNA melting temperature below room temperature. Upon denaturing, the FDCD signal reverses sign or is destroyed completely (Figures 3c, S4). Importantly, because the dye-functionalized ssDNA remains tethered to the surfaces, the change in FDCD signal can be attributed to partial or complete dehybridization of adsorbed DNA rather than removal of fluorescent probes from the surfaces.
\begin{figure}
\includegraphics{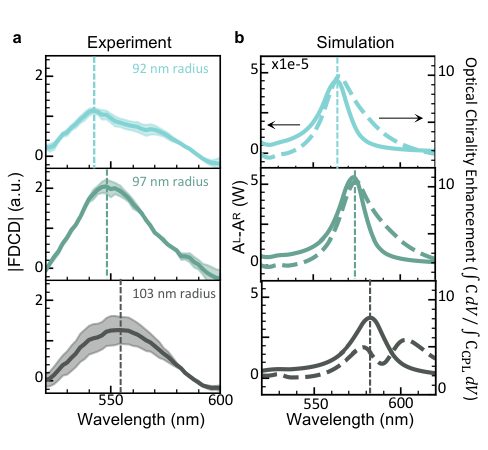}
\caption{\label{fig4}a) Normalized FDCD signal from dye-dsDNA on $r$=92-103nm arrays (details in SI). Vertical lines mark spectra maxima, illustrating red-shift, shaded region represents standard error. b) Simulated $\Delta$A from 10nm chiral layer on disk surface (left axis, solid) against integrated $\text{C/C}_{\text{CPL}}$ (right axis, dotted). Vertical lines mark $\Delta$A maxima.}
\end{figure}

We repeat the double-stranded DNA measurements on the three metasurfaces characterized in Figure 1 to study the spectral influence of the metasurface resonance on the FDCD signal. These results (Figure 4a) show a red-shifting FDCD signal with increasing disk size. Consistent with $\text{C/C}_{\text{CPL}}$ in Figure 1, the peak is blue-shifted from the resonance center wavelength. In contrast, upon repeating the measurement on an identically-sized square of unpatterned silicon on the same sapphire substrate, no significant FDCD signal is observed (Figure S5), illustrating that the detected signal can primarily be attributed to the electromagnetic near-field enhancements. To confirm that the measured FDCD signal arises primarily from molecular circular dichroism, we perform full-field simulations, calculating $\Delta$A and $\text{C/C}_{\text{CPL}}$ within a 10-nm-thick chiral layer above the silicon disks. To account for coupling effects between induced electric and magnetic dipoles in the chiral medium, these simulations use a wavelength-independent Pasteur parameter with strength typical of an on-resonant chiral molecule (see SI).  \cite{garcia2018enantiomer,mohammadi2019accessible,nesterov2016role} Figure 4b shows these results, indicating that the peak in molecular $\Delta$A occurs just blue-shifted from the dip in transmission, and red shifts with increasing disk size, corresponding well with experiments. Further, the observed lineshape and spectral position of the simulated $\Delta$A are in good agreement with those of the optical chirality enhancement in the molecular layer, seen in the solid and dotted lines of Figure 4b, respectively. We note that the optical chirality calculated in the chiral layer on the 103-nm-metasurface exhibits two peaks, in contrast to the single peak in $\Delta$A, likely due to higher enhancements in electric field associated with the blue-shifted mode as the resonances separate. Importantly, $\Delta$A within the silicon nanodisks in simulation (Figure S6) is given by a bisignate lineshape not seen in our measurement. This result further indicates that the measured FDCD signal arises primarily from enhanced molecular CD due to high optical chirality density in the near field, as opposed to induced CD in the disks themselves.\cite{garcia2019enhanced} We attribute the non-negligible ($\sim$~20nm) red shift between simulation and experiment to a combination of an oxide layer on the silicon surface, tapering of disk side-walls, and wavelength-dependent chiral medium properties which were not captured in our model. 
	
In summary, we demonstrate enhancement of intrinsic molecular CD in a monolayer via overlap of molecular CD with nanophotonic resonances using FDCD. We study the circular dichroism of $\text{Atto}^{\text{TM}}$590-functionalized DNA strands, with a solution-phase CD signal from 500-650nm. Interestingly, the sign of this CD signal is dependent on DNA conformation, with dye-dsDNA exhibiting a positive signal and dye-ssDNA exhibiting a negative signal. The complexes are functionalized to silicon metasurfaces fabricated with optical resonances in the CD band of interest and characterized using a table-top FDCD spectrometer. We exploit the enantiomer-reversal-like behavior of the different conformations to probe surface chirality by de-hybridizing the dsDNA, observing a sign-reversal when metasurfaces are present but negligible signal in their absence. Finally, we show that the FDCD signal red shifts with increasing disk size, and thus that the metasurface's resonant features are central to molecular CD enhancement. Full-field simulations confirm that both differential absorption and optical chirality density are enhanced within the chiral medium near the disk resonances. This indicates that the experimental FDCD enhancement is due to enhanced optical chirality density, as predicted in prior works, thus providing important insights toward efficient enantioselective photolysis.

\section{Acknowledgements}

The authors thank Dr. Mark Lawrence, Dr. Aitzol Garc{\'i}a-Etxarri, Dr. Michal Vadai, Dr. Claire McClellan, and David Barton for insightful discussions. The authors gratefully acknowledge the Gordon and Betty Moore Foundation for funding through a Moore Inventors Fellowship under grant number 6881, and the National Science Foundation under grant number 1905209. M.L.S. acknowledges a National Defense Science and Engineering graduate fellowship. L.V.P. acknowledges support from a Swiss National Science Foundation Early Postdoc Mobility fellowship under project number P2EZP2\_181595. Work was performed in part in the nano$@$Stanford labs (Stanford Nano Shared Facilities and the Stanford Nanoafabrication Facility), which are supported by the National Science Foundation as part of the National Nanotechnology Coordinated Infrastructure under award ECCS-1542152.

\subsection{Supporting Information}

Supporting Information contains materials and methods, a summary of findings of dye-DNA functionalized to unpatterned silicon, simulation details, and raw data and data analysis procedure, as well as additional data.

\bibliography{refs}

\end{document}

%\begin{titlepage}
%\tableofcontents
%\listoffigures
%\end{titlepage}
\section{Background on Mie Resonances and Optical Chirality Enhancement}

\begin{figure}[ht]
\includegraphics{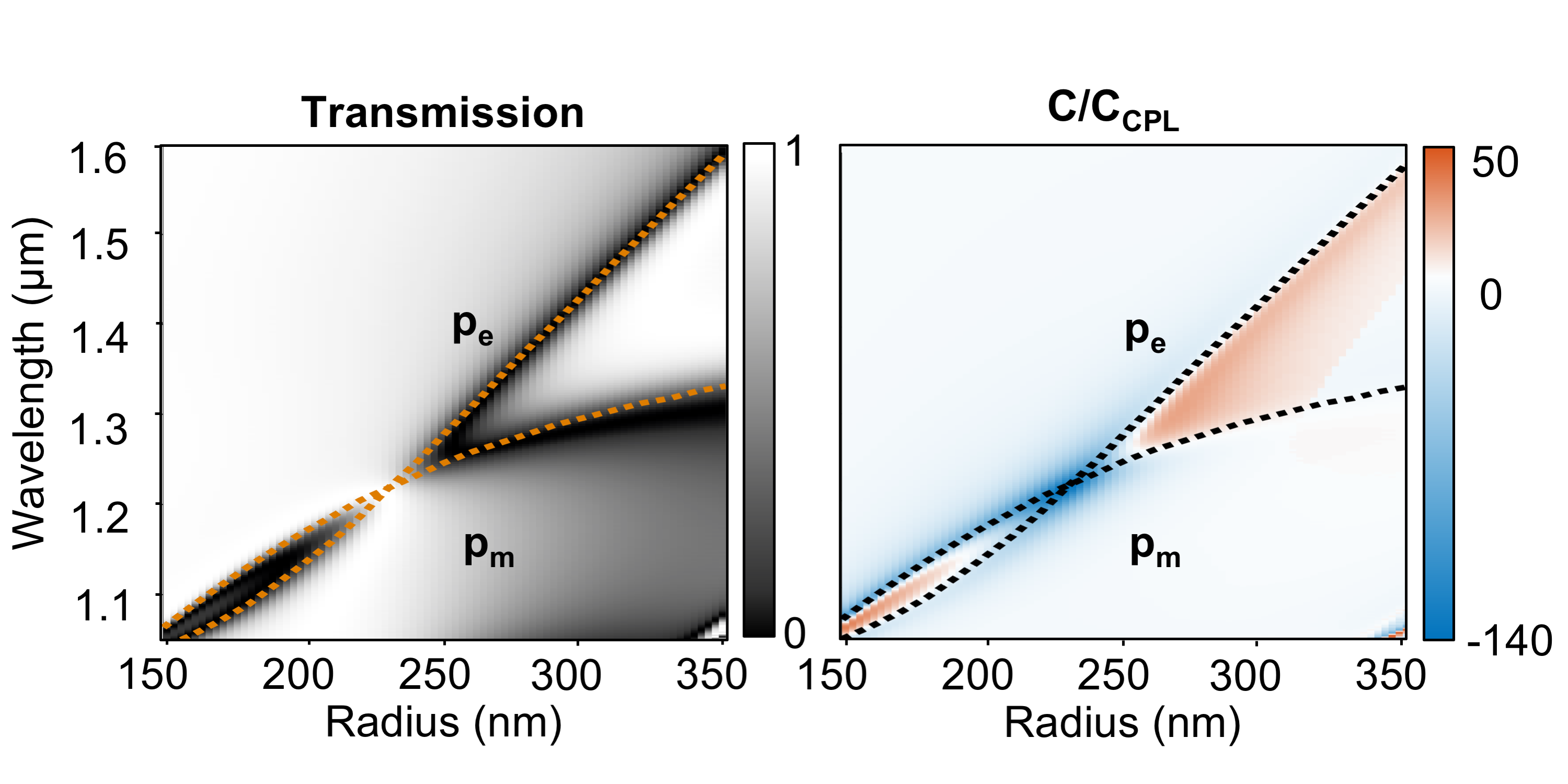}
\caption{\label{EDMD} Relationship between electric and magnetic dipole resonance and optical chirality enhancement. This figure has been adapted with permission from Solomon et al, Copyright American Chemical Society 2018\cite{solomon2018enantiospecific} and Solomon et al, Copyright American Chemical Society 2020\cite{solomon2020nanophotonic}}
\end{figure}

The relationship between electric and magnetic dipoles and enhancement in optical chirality density has been thoroughly studied in our previous work, in which we optimized disk aspect ratio for enhancements in optical chirality. As seen in Figure S\ref{EDMD}, we found that as the radius of the disk increases while the height and spacing is kept constant, the electric resonance red-shifts faster than the magnetic resonance and the two come to spectrally overlap before separating again. We saw that for disks that are on the blue side of that overlap (with a smaller radius compared to the height), the optical chirality enhancements are stronger near the magnetic dipole mode, but on the red side of the overlap (with a larger radius compared to the height), the optical chirality enhancements are stronger on the electric dipole resonance. However, the overall maximum enhancement occurs when the two resonances are centered on the same wavelength. At this point, both the electric and magnetic field are strong, and because the resonances have the same center wavelength, the phase lag between the fields maintains that of the incident light, which in this case is circularly polarized light. As described in the main text, optical chirality density is equal to $\text{C}=-\frac{\omega}{2}\text{Im}(\textbf{D}^*\cdot\textbf{B})$, which is equivalent to $\text{C}=-\frac{\omega}{2}|\textbf{D}||\textbf{B}|\text{cos}(\beta_{i\textbf{D},\textbf{B}})$. Thus, the optical chirality density is proportional to the strength of the electric and magnetic fields, in addition to the cosine of the phase lag between $i$\textbf{D} and \textbf{B}. For perfectly circular light, $\text{cos}(\beta_{i\textbf{D},\textbf{B}})=1$, giving a maximum in optical chirality density. Therefore, the largest enhancements occur when electric and magnetic fields are strong, and the phase lag between them is that of CPL, just as they are when the electric and magnetic dipole are overlapped. 
\section{Circular Dichroism of DNA-dye Complex}

\begin{figure}
    \centering
    \includegraphics[width=10cm]{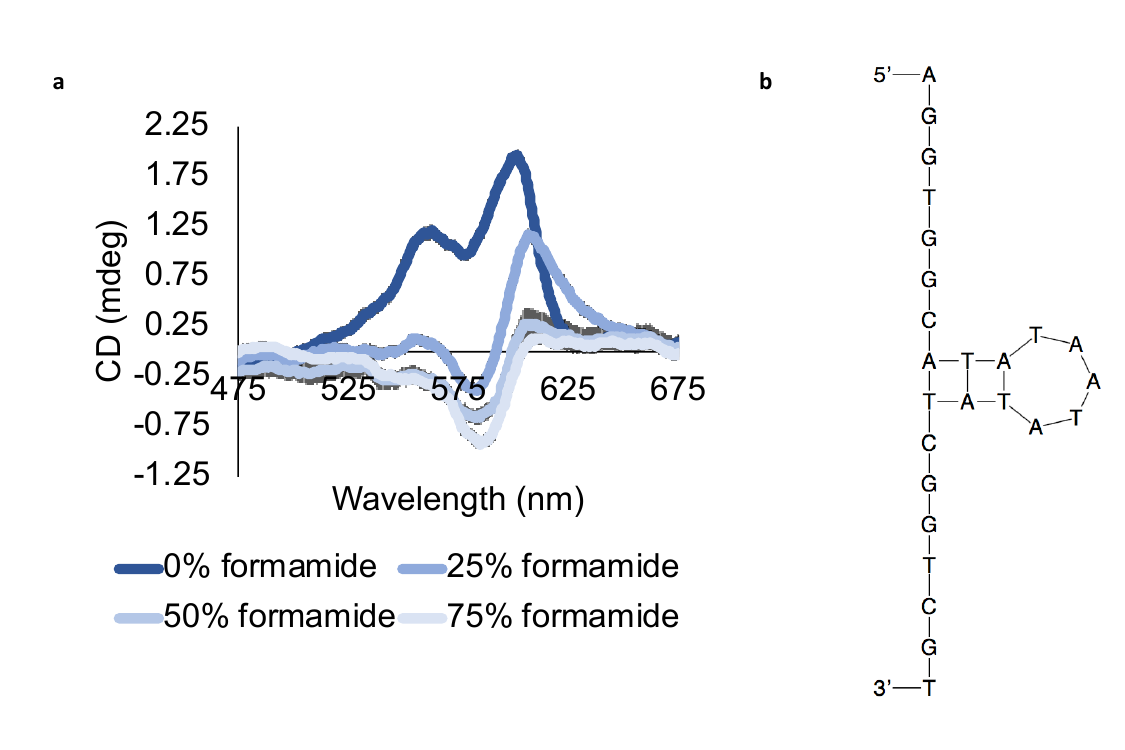}
    \caption{a) Circular dichroism of ATTO 590 functionalized to dsDNA, changing depending on conformational changes due to addition of formamide b) Likely stable conformation of single stranded DNA used in this work.}
    \label{ssDNA}
\end{figure}

The selection of the DNA strands used in this work was based on previous literature reporting CD signals of dye-DNA complexes.\cite{takada2017luminescent} We performed CD measurements for the dye-functionalized DNA strands at various fractions of single vs double-stranded conformations using different concentrations of formamide in PBS solution (Figure S\ref{ssDNA}a). The 100\% formamide solution is expected to de-hybridize the DNA completely from double- to single-stranded, with partial de-hybridization occurring at intermediate concentrations, resulting in changing CD signals as expected based on measurements in previous literature of dyes sitting within minor grooves of DNA.\cite{stokes2007making,norden1982structure} Figure 3 in the main text plots the complete double- and single-stranded conformations. We have estimated the structure of our single-stranded oligomer using the open source software "mfold" to predict the secondary structure of the single strand based on a PBS solution.\cite{zuker2003mfold} The most stable conformation is shown in Figure S\ref{ssDNA}b.

\section{Materials}
All chemicals were purchased from Sigma-Aldrich Co.(St. Louis, MO). Oligonucleotides were obtained from Integrated DNA Technologies
(Coralville, IA). Silicon-on-sapphire substrates were purchased from MTI Corp (Richmond, CA). 
\section{Methods}

\subsection{Fabrication}
Diced crystalline silicon-on-sapphire wafers were purchased from MTI corp. The initial height of the silicon layer was 600 nm, so a pre-etch using reactive ion etching in Lam Research 9400 TCP etcher with 100 sccm C$_2$F$_6$ for 10s to break through the native oxide, and then 40 sccm Cl$_2$, 100 sccm HBr, and 5.1 sccm 20\% O$_2$ in He was performed to reach a height of 80 nm, which was confirmed using ellipsometry. After solvent cleaning in acetone, methanol, and isopropanol, we bake the substrate for 2 minutes at 180\degree C. After cooling, the sample is then dipped in Surpass 4000 for 30 seconds for passivation. The sample is rinsed in water and dried with nitrogen. A nominally 100 nm layer of ma-N 2401 resist is spun on at 3000 rpm for 60s, and the sample is then baked for 90s at 90\degree C on a hot plate. We then coat a layer of the conductive polymer e-spacer, spinning it on at 2000 rpm. We pattern 100 by 100 $\mu$m arrays of circles using a JEOL JBX 6300 electron beam lithography system. The samples are then dipped in water to remove the e-spacer, developed for 35 s in MF 319 developer solution, and rinsed twice in water. The samples were then etched using the same reactive ion etching system with 100 sccm $\text{C}_2\text{F}_6$ for 10s to break through the native oxide, and 40 sccm $\text{Cl}_2$, 100 sccm HBr, and 5.1 sccm 20\% $\text{O}_{2}$ for 25 s, leaving behind only the exposed pattern. The rest of the resist is cleaned from the sample in an an acid piranha clean.

\subsection{Optical Measurements}
\begin{figure}[ht]
\includegraphics[width=\textwidth]{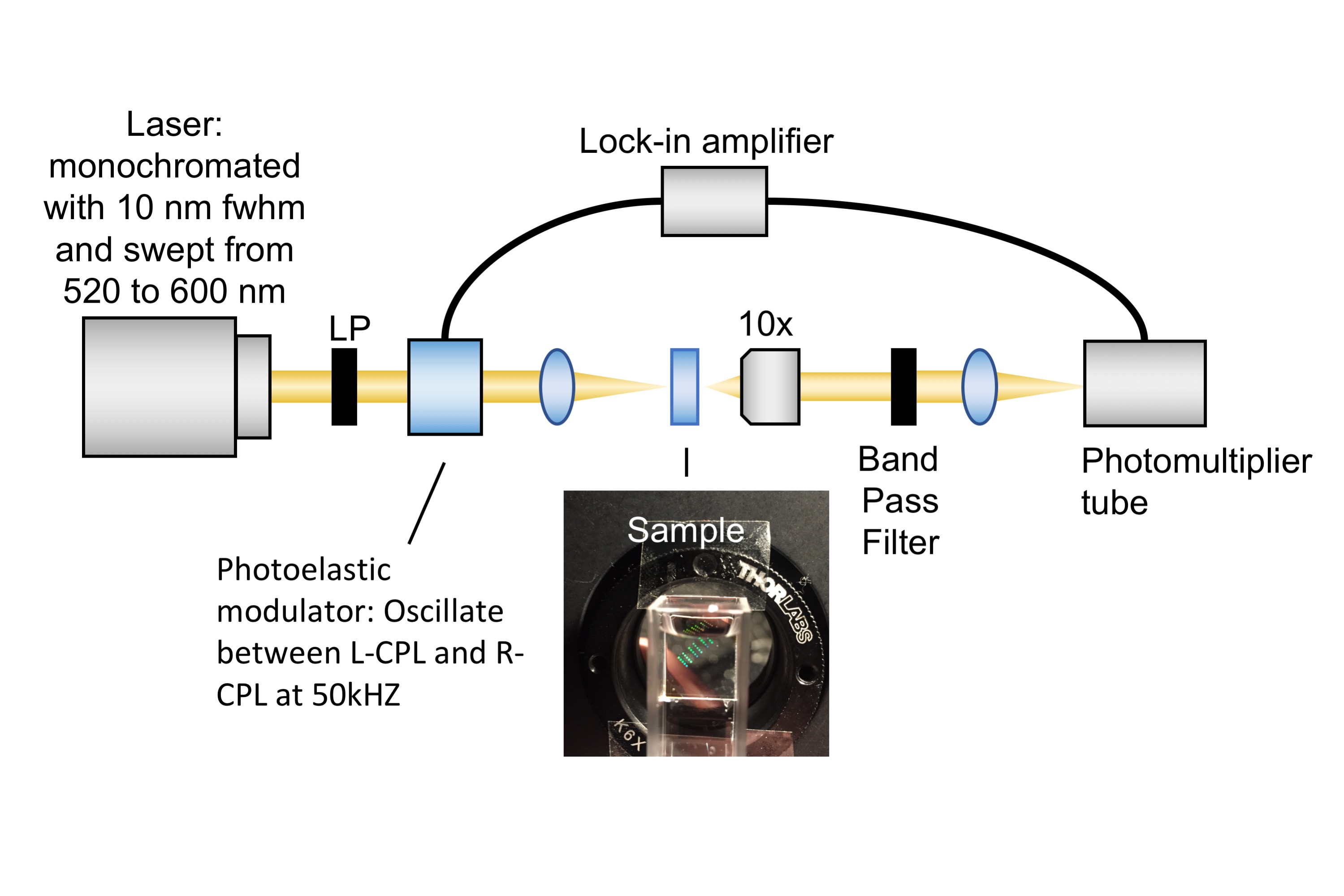}
\caption{\label{figoptical}Schematic of table-top fluorescence detected circular dichroism spectrometer}
\end{figure}

A table-top microscope was constructed using an NKT Supercontinuum Extreme laser with a Varia filter, photoelastic modulator (Hinds Instruments, I/FS50), 10x objective lens with a 0.3 NA and 10 mm working distance (Olympus, RMS10X-PF), and photomultiplier tube (PMTSS, Thorlabs), in addition to a a 635 nm bandpass filter with an 18 nm bandwidth (Semrock, FF01-635/18-25), see Figure S\ref{figoptical}. We sweep the excitation center wavelength from 520 nm to 600 nm using the Varia filter (which has a 10 nm FWHM bandwidth at each center wavelength). The light first passes through a linear polarizer, and then the photoelastic modulator (PEM, Hinds Instruments, I/FS50), resulting in R- and L-CPL excitation oscillating at 50 kHz. This oscillation is fed into a lock-in amplifier (Stanford Instruments, SR865) as the input frequency. The light then passes through the sample (substrate side first), and is collected using a 10x objective. We focus on the sample using an alignment camera (Thorcam), and the laser spot is focused down to approximately 100 $\mu$m in diameter, roughly the size of our metasurface arrays. After filtering out the excitation light using the bandpass filter, and focusing into the photomultiplier tube using a lens, we measure the total fluorescence intensity. The signal from the photomultiplier tube, modulated at 50 kHz by the PEM, is then fed into the voltage channel on the lock-in amplifier. The amplitude of the oscillation ($V_{RMS}$) of the detected light as well as the phase lag from that input oscillation ($\theta$), is recorded at each wavelength. The circular dichroism value at each wavelength is calculated by:
 
\begin{equation}
    \text{CD}=-V_{RMS}*\text{sin}(\theta)
\end{equation}
 
Here, we multiply by -1 to account for the fact that we are doing a fluorescence measurement, not an absorption measurement as in a conventional CD spectrometer. We also normalize by the laser power by performing a non-polarization modulated lock-in measurement using a chopper. In this experiment, a chopper is inserted in the beam path before the linear polarizer and PEM, at 500 Hz rotation frequency. This oscillation becomes the input of the lock-in amplifier, and we record the amplitude of the oscillation in the fluorescence signal as $V_{DC}$. The CD signal normalized by laser power is then:
 
\begin{equation}
    \text{CD}_{norm}=-\frac{V_{RMS}*\text{sin}(\theta)}{V_{DC}}
\end{equation}
 
We operate the laser at 100\% power to ensure that the power is as consistent as possible across the spectrum of interest, while neutral density filters at the source attenuate the light and prevent oversaturation of the dye and detector. The sample is mounted inside of a single-use cuvette using a line of vacuum grease along the bottom of the sample, and the cuvette is filled with PBS. We compensate for the intrinsic birefringence of the sapphire substrate by inserting an additional sapphire substrate, where the silicon layer is completely etched, into the beam path, rotated at 90 degrees and preceding the sample of interest. The measurements were performed wavelength by wavelength, with the lock-in time constant set to 100 ms, and the measured value being an average of five points. 

\subsection{Self-Assembled Monolayer Functionalization}

First, our patterned substrates were cleaned in 9:1 H$_2$SO$_4$:H$_2$O$_2$ solution at 120\degree C for 20 minutes. Then, (3-aminopropyl)trimethoxysilane (APTMS) was vapor deposited in samples at 40\degree C for 1 hr by thermal evaporation of neat solutions. Next, a solution of 1 mM \textit{m}-maleimidobenzoyl-\textit{N}-hydroxysuccinimide ester (MBS) in 1:9 DMSO:PBS(1x) was mixed. The substrate was baked on a hotplate at 40\degree C for 10 minutes, and then submerged in the MBS in DMSO:PBS solution for 1 hour. The substrate was rinsed in ethanol and water, and dried with argon. Finally, a hybridized solution of 10 $\mu$M DNA and dye in 1x PBS was incubated on the surface in a dark, humid environment for 24-48 hours, using 20-40 $\mu$L of solution on each sample.\cite{nakatsuka2018aptamer} 

To prepare the DNA for functionalization, desalted and lyophilized single-stranded 25 base pair oligonucleotides, as received from IDT, containing a disulfide tether on the 3' end and ATTO$^{\text{TM}}$ 590 dye on the 5'-end (5'-AG GTG GCA TAT AAT ATA TCG GTC GT-3') were dispersed in 50 $\mu$L tris-EDTA buffer, pH 8.0, and mixed with $\sim$30 mg of DL-dithiothreitol for at least 1 h to reduce the disulfide moieties to thiols. Oligonucleotides were then purified via gravity-flow size exclusion chromatography using illustra NAP-5 columns. The eluent DNA concentrations were determined using UV absorption signatures (Varian Cary 5000 UV-Vis Spectrophotometer). Desalted and lyophilized single-stranded oligonucleotides, as received from IDT, that did not contain disulfide or dye modifications and were the complement of those that did (5'-ACG ACC GAT ATA TTA TAT GCC ACC T-3'), were simply dispersed in tris-EDTA buffer, pH 8.0, without further purification. To anneal oligonucleotides and to create double-stranded DNA/dye complexes for surface assembly, complementary strands of stock DNA and thiol/dye functionalized DNA were combined in 1x PBS to attain 10 $\mu$M DNA concentration. Solutions were bubbled with Ar and annealed at 95\degree C for 5 min, followed by slow cooling at room temperature. An appropriate volume of 1 M MgCl$_2$ was added before incubation on surfaces for a final concentration of 100 mM MgCl$_2$ to minimize electrostatic repulsion between negatively charged backbones of oligonucleotides and to form denser DNA assemblies on surfaces.\cite{cheung2020detecting}

\subsection{Non-Polarization Sensitive Fluorescence Measurements}

Fluorescence measurements were performed in a Zeiss inverted microscope (Zeiss, Axio Observer) using a 10x objective (Zeiss, 0.2 NA, bright-field dark-field, air, EC EPIPLAN (422040-9960)) and a 550 nm bandpass-filtered halogen source (Zeiss, HAL). Light was routed through a 600 nm long pass filter to collect emission and through a spectrometer (Princeton Instruments, Acton SP2500) to a liquid-nitrogen cooled CCD camera (Princeton Instruments, PyLoN:400BR eXcelon). 

\subsection{Solution Phase Measurements of dye-DNA}
Solution phase circular dichroism measurements of the dye-DNA complexes were done in a Jasco J-815 CD spectrometer. Solution phase excitation and emission measurements were done in a Varian Cary Eclipse Fluorescence Spectrophotometer. Excitation spectra were collected by measuring emission strength at 635 nm, and emission spectra were collected when exciting 500 nm, 525 nm, 550 nm, 575 nm, and 600 nm.

\subsection{Denaturing}

To perform denatured CD measurements, we use formamide to de-hybridize the DNA from double to single stranded. We do this by replacing the PBS with formamide in the cuvette, and letting the system equilibrate for 15 minutes. At this point, the DNA strand with no dye or thiol is released from the substrate and becomes extremely dilute in solution. The strand with the thiol modification and the dye remains attached, as depicted in Figure 3a in the main text. 
\begin{figure}[h!]
\includegraphics{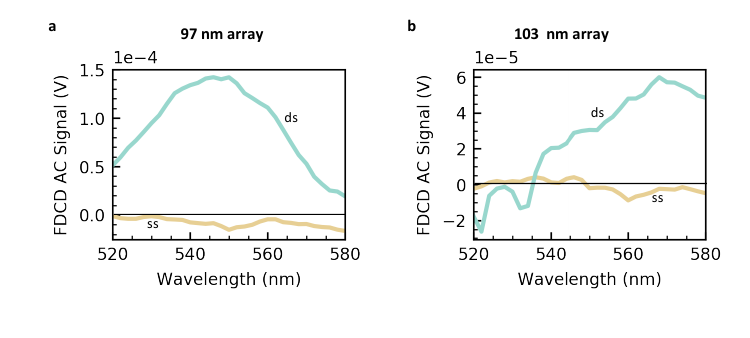} 
\caption{\label{figdenat} a) 97 nm array and b) 103 nm array FDCD AC signal showing sign change in FDCD from double to single stranded.}
\end{figure}

We also measure the results of denaturing on the 97 nm array and 103 nm array, finding a sign change in the FDCD signal in both cases (Figure S\ref{figdenat}), matching Figure 3c in the main text. The signal for the ssDNA is much lower for these two arrays than for the 92 nm array shown in the main text, possibly due to incomplete de-hybridization.

\section{Summary of Findings on Flat Silicon}
\begin{figure}[ht]
\includegraphics{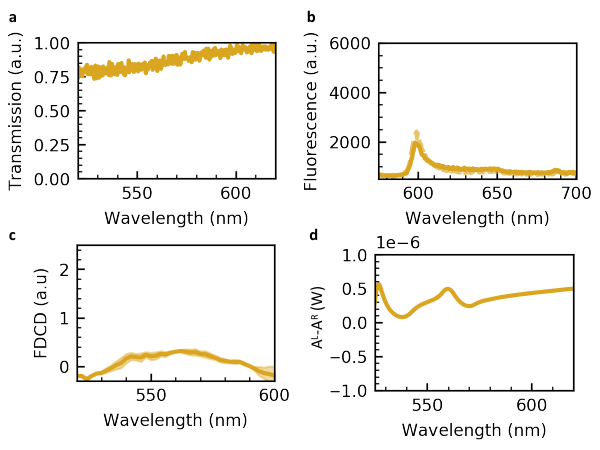}
\caption{\label{figflat}a) Transmission of flat silicon square on sapphire substrate in PBS b) Flat silicon square with and without dye functionalized, showing no significant enhancement in fluorescence without the nanodisks. c) Experimental FDCD signal on flat silicon square, showing no significant peak in FDCD d) Simulated differential absorptance of chiral layer on top of flat silicon, 2 orders of magnitude lower than the differential absorptance of the chiral layer on top of the nanodisks shown in the main text.}
\end{figure}
On the flat silicon square of identical size to the nanopatterned arrays (100 $\mu$m by 100 $\mu$m), we see features that are very different from those on the arrays. First, in transmission, we see no dips, which means that there are no resonances present and thus no enhanced near fields. When we functionalize the flat silicon square with dye-DNA, we see no fluorescence, indicating that not only was the dye present on the disk arrays, but also that the nanodisks were enhancing that fluorescence. When we look at the FDCD, we also see no significant peak in the signal, further supporting the fact that the near-fields of on-resonant structures are crucial to the FDCD signals shown in the main text. Finally, the differential absorption in the chiral layer on top of flat silicon in simulation is two orders of magnitude lower than the enhanced differential absorption seen in the layers on top of the disks in the main text.

\section{Simulations}

\begin{figure}[ht]
\includegraphics{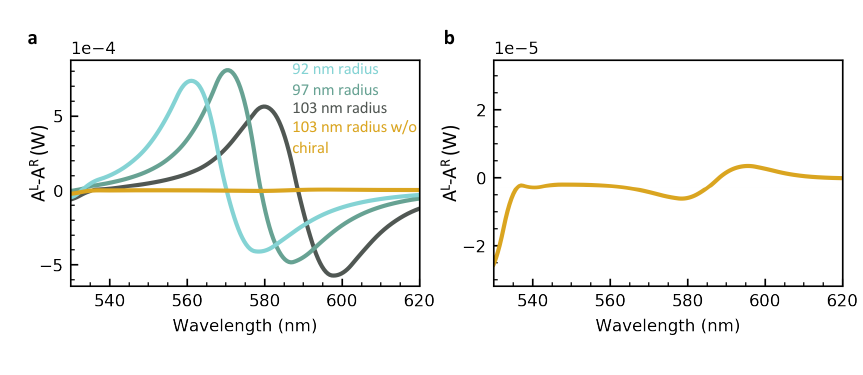}
\caption{\label{figsidiff}a) Differential absorptance in silicon for all three disk arrays when chiral medium is present showing induced CD bisignate lineshape, and with no chiral medium present on the 103 nm disk array b) Differential absorptance in 103 nm silicon disk with no chiral medium present, on a finer y-axis scale showing that there is a background CD due to imperfect meshing, but it is significantly lower than that when the chiral medium is present, or that within the chiral medium (shown in the main text).}
\end{figure}

The numerical simulations done in this work use the finite element method via COMSOL Multiphysics version 5.3, resulting in time-harmonic, electromagnetic near fields
governed by Maxwell’s equations. Data from Aspnes were used for the optical
material constants of the silicon nanodisks.\cite{aspnes1979dielectric} A silicon nanodisk was positioned with its center at the (0,0,0) location in the simulation region. It was positioned on a substrate ($n_{substrate} = 1.77$, that of sapphire) occupying the entire half space below the nanoantenna, with PBS ($n_{PBS} = 1.335$) occupying the half space above. Circularly polarized plane-wave excitation (x-polarization = 1, y-polarization  = +/- i, propagating in +z direction) was incident from a port at the lower domain boundary (z = -840 nm plane). We simulate a region 300 nm in the x- and y-directions and 1680 nm in the z-direction, with Bloch boundary conditions in the x- and y-directions and ports in the z-directions. The input power is 1 W through the port. To collect the differential absorptance, in the chiral medium and silicon layers, we calculate the absorptance under left and right-CPL illumination as:

\begin{equation}
    A=2\omega\int\text{Im}(\textbf{E}\cdot \textbf{D}^*-\textbf{B} \cdot \textbf{H}^*)dV,
\end{equation} 

and integrate over either the chiral medium or silicon domains. We calculate the optical chirality density as:

\begin{equation}
C=-\frac{\omega}{2}(\textbf{D}^* \cdot \textbf{B}).
\end{equation}

To find the enhancements in optical chirality and the electric field, we run a simulation with no nanostructure or substrate, and use the background values, which in this case were found to be: $|\textbf{E}|=7.92*10^{7}$ V/m, and $C=7.345*10^{11}$ $\text{N}/\text{m}^{3}$.

To simulate the chiral medium, we position a a thin disk layer 10 nm in height, centered on top of the disk, with radius 10 nm smaller than the silicon disk (82 nm, 87 nm, and 93 nm for the 92 nm silicon disk, 97 nm silicon disk, and 103 nm silicon disk respectively). Within this region, we modify the constitutive relations to account for coupling effects between electric and magnetic dipoles in chiral media. These constitutive relations that define chiral media and are used in this work are:\cite{garcia2018enantiomer,nesterov2016role,mohammadi2019accessible}  

\begin{equation}
    \textbf{D}=\epsilon\textbf{E}-\frac{i\chi}{c_0}\textbf{H}
\end{equation}
\begin{equation}
    \textbf{B}=\epsilon\textbf{H}-\frac{i\chi}{c_0}\textbf{E}
\end{equation}

The chiral medium we consider has a Pasteur paramater, $\chi$, with the strength of an on-resonant chiral molecule (Re($\chi$)=$7*10^{-4}$, and Im($\chi$)=$1*10^{-6}$). It has a refractive index of 1.335 to match that of the PBS it is submersed in, with losses (imaginary refractive index) of 0.01 to account for the strong absorption of the dye. In COMSOL syntax, the constitutive relation are expressed for D, H, and dH/dt as follows:
\newline
For D:
\newline
emw.Dx=epsilon0\_const*emw.Ex+emw.Px-i/c\_const*$\chi$*emw.Hx
\newline
emw.Dy=epsilon0\_const*emw.Ey+emw.Py-i/c\_const*$\chi$*emw.Hy
\newline
emw.Dz=epsilon0\_const*emw.Ez+emw.Pz-i/c\_const*$\chi$*emw.Hz.
\newline
\newline
For H:
\newline
emw.Hx=(emw.murinvxx*emw.Bx+emw.murinvxy*emw.By
\newline
+emw.murinvxz*emw.Bz-i/c\_const*$\chi$*(emw.murinvxx*emw.Ex
\newline
+emw.murinvxy*emw.Ex+emw.murinvxz*emw.Ez))/mu0\_const
\newline
\newline
emw.Hy=(emw.murinvyx*emw.Bx+emw.murinvyy*emw.By
\newline
+emw.murinvyz*emw.Bz-i/c\_const*$\chi$*(emw.murinvyx*emw.Ex
\newline
+emw.murinvyy*emw.Ey+emw.murinvyz*emw.Ez))/mu0\_const
\newline
\newline
emw.Hz=(emw.murinvzx*emw.Bx+emw.murinvzy*emw.By
\newline
+emw.murinvzz*emw.Bz-i/c\_const*$\chi$*(emw.murinvzx*emw.Ex
\newline
+emw.murinvzy*emw.Ey+emw.murinvzz*emw.Ez))/mu0\_const.
\newline
\newline
For dH/dt:
\newline
emw.dHdtx=(emw.murinvxx*emw.dBdtx+emw.murinvxy*emw.dBdty
\newline
+emw.murinvxz*emw.dBdtz+emw.omega/c\_const*$\chi$*(emw.murinvxx*emw.Ex
\newline
+emw.murinvxy*emw.Ey+emw.murinvxz*emw.Ez))/mu0\_const
\newline
\newline
emw.dHdty=(emw.murinvyx*emw.dBdtx+emw.murinvyy*emw.dBdty
\newline
+emw.murinvyz*emw.dBdtz+emw.omega/c\_const*$\chi$*(emw.murinvyx*emw.Ex
\newline
+emw.murinvyy*emw.Ey+emw.murinvyz*emw.Ez))/mu0\_const
\newline
\newline
emw.dHdtz=(emw.murinvzx*emw.dBdtx+emw.murinvzy*emw.dBdty
\newline
+emw.murinvzz*emw.dBdtz+emw.omega/c\_const*$\chi$*(emw.murinvzx*emw.Ex
\newline
+emw.murinvzy*emw.Ey+emw.murinvzz*emw.Ez))/mu0\_const.

\begin{figure}[ht]
\includegraphics[width=8 cm]{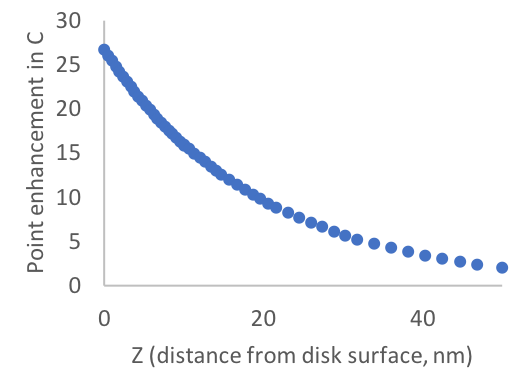}
\caption{\label{figz} Point enhancement in optical chirality above the center of the disk.}
\end{figure}

\section{Raw Data and Data Analysis}
\begin{figure}[ht]
\includegraphics{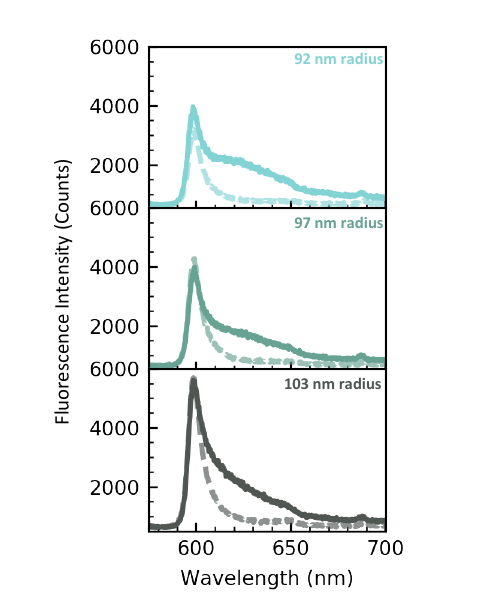}
\caption{\label{figfluor}Fluorescence on 92, 97, and 103 nm arrays with (solid lines) and without (dotted lines) dye. }
\end{figure}

In addition to the dye fluorescence on the 92 nm array, shown in the main text, we also measure fluorescence with and without dye on the 97 nm array and the 103 nm array, confirming that fluorescence is present in each of these instances, as well (Figure S\ref{figfluor}). 

\begin{figure}[ht]
\includegraphics[width=\textwidth]{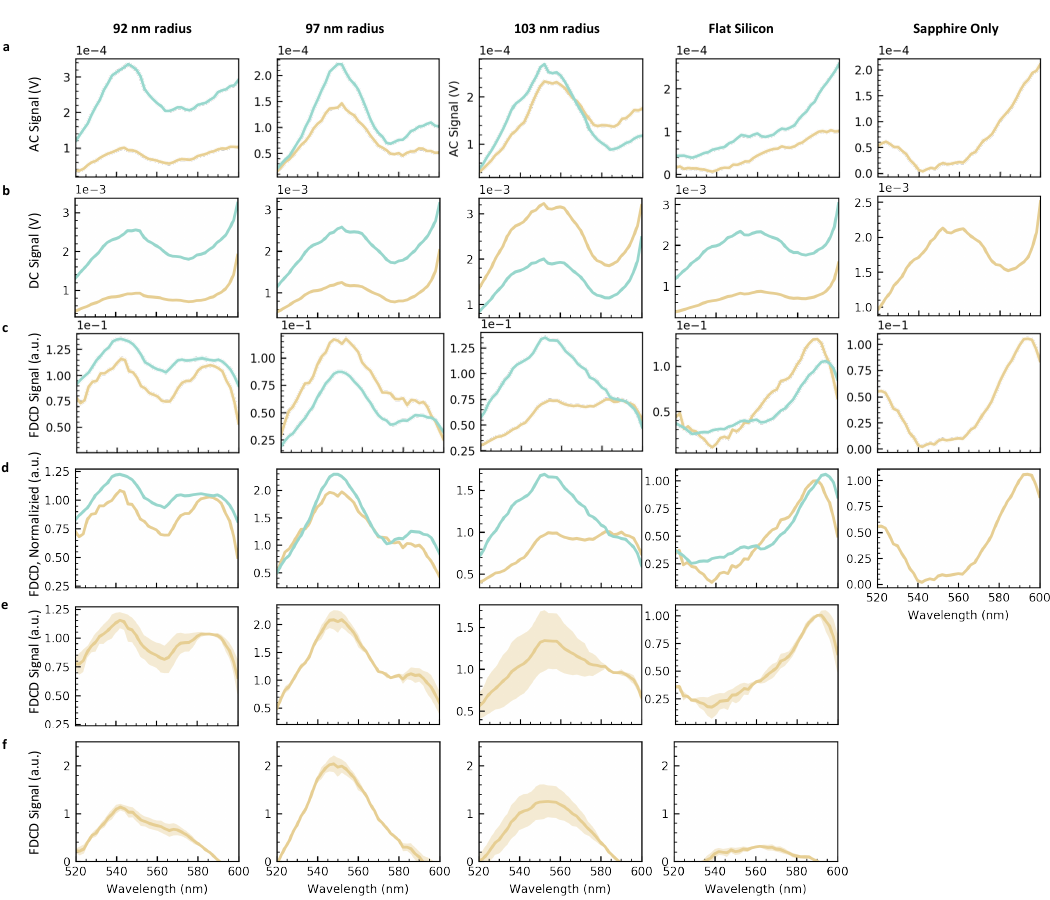}
\caption{\label{figlast} Workflow of data analysis from raw data to data in main text a) AC Signal as calculated by $-V_{RMS}*\text{sin}(\theta)$ b) DC signal, $V_{DC}$ c) FDCD signal as calculated by $-V_{RMS}*\text{sin}(\theta)/V_{DC}$ d) FDCD signal normalized by value of spectrum at 590 nm e) Averaged FDCD spectrum plotted with standard error f) Averaged spectrum with sapphire background subtracted.}
\end{figure}

To calculate the circular dichroism signal, we measure a polarization modulated signal at 50 kHz, recording the amplitude and phase of this oscillation. From this signal, we extract the $V_{RMS}$ and the phase of the oscillation, and multiply them together to get an AC signal, shown Figure S\ref{figlast}a. Additionally, we measure a non-polarization modulated signal using an optical chopper, rotating at 500 Hz. This non-polarization modulated signal serves as the normalization of laser power, and is shown as the DC signal in Figure S\ref{figlast}b. From these two values, we calculate FDCD=$V_{RMS}*\text{sin}(\theta)/V_{DC}$, or $V_{AC}/V_{DC}$. For the FDCD signal in Figure S\ref{figlast}c for all samples, we see a peak occurring at 590 nm due to the increase in DC voltage due to the edge of the bandpass filter we use. Therefore, we normalize all samples to their maxima at 590 nm to account for any variations in how the laser spot was incident on the arrays. The results of this normalization are seen Figure S\ref{figlast}d. For the three arrays and the flat silicon square, we took two measurements, performed after leaving the array and coming back to ensure that the signal from each array was repeatable and not dependent simply on the position of the laser spot in that moment. We then averaged these two measurements, and took the standard error, as shown in Figure S\ref{figlast}e. Finally, in each of these plots, we still see an upwards slope toward  and peak at 590 nm, including on the sapphire substrate, where there should be very little dye bound based off of the binding efficiency to sapphire vs silicon. Therefore, we conclude that this upward slope and peak are due to background birefringence in this round of measurements, and we subtract the background from the three disk arrays and Figure S\ref{figlast}f and the main text Figure 4a. 

\bibliography{refs}